\documentclass[twocolumn,prl]{revtex4}
\usepackage{graphicx}    
\usepackage{dcolumn}     
\usepackage{bm}          
\newcommand{\Vec}[1]{\mbox{\boldmath$#1$}}
\begin{document}
\title{ ``Pudding mold'' band drives large thermopower in Na$_x$CoO$_2$}
\author{
Kazuhiko Kuroki$^1$ and Ryotaro Arita$^2$ 
}
\affiliation{
$^1$ Department of Applied Physics and Chemistry, 
The University of Electro -Communications, Chofu, Tokyo 182-8585, Japan\\  
$^2$ RIKEN, 2-1 Hirosawa, Wako, Saitama 351-0198, Japan
}
\date{\today}
\begin{abstract}
In the present study, we pin down the origin of 
the coexistence of the large thermopower 
and the large conductivity in Na$_x$CoO$_2$.
It is revealed that not just 
the density of states (DOS), the effective mass, 
nor the band width, but 
the peculiar {\it shape} of the $a_{1g}$ band referred to as the 
``pudding mold'' type, which consists of  
a dispersive portion and a somewhat flat portion, 
is playing an important role in this phenomenon.
The present study provides a new guiding 
principle for designing good thermoelectric materials.
\end{abstract}
\pacs{PACS numbers: }
\maketitle

For the past decade, a cobaltate Na$_x$CoO$_2$ 
has been one of the most highlighted materials 
in the field of condensed matter 
physics in that it exhibits large thermopower\cite{Terasaki} and 
magnetism\cite{Motohashi} in the non-hydrated sodium rich systems, 
and superconductivity in the hydrated sodium poor ones.\cite{Takada} 
In particular, the discovery of large thermopower in Na$_x$CoO$_2$
\cite{Terasaki} and the findings in  
cobaltates/cobaltites\cite{Li,Fujita,Hebert,Kajitani,Lee} 
and rhodates\cite{Okada,Takagi} that followed 
have brought up an interesting possibility of 
finding good thermoelectric materials that have 
relatively large conductivity. 

Theoretically, it has been proposed that the three fold degeneracy of the 
$t_{2g}$ orbitals as well as the 
strong electron correlation effects 
plays an important role in these transition metal oxides.\cite{Koshibae}
Importance of spin degeneracy has been pointed out from 
experiments under magnetic field.\cite{Wang}
As for the orbital degeneracy, 
the $t_{2g}$ orbitals are split into 
an $a_{1g}$ orbital and two $e_g'$ orbitals due to the crystal field,
and early first principles calculation 
predicted six hole pockets from the $e_g'$ bands
in addition to the $a_{1g}$ Fermi surface.\cite{Singh}
However, angle resolved photoemission spectroscopy (ARPES) studies 
\cite{Hasan,Yang2,Takeuchi,Shimojima}
have revealed that the $e_g'$ bands 
lie $O(100)$meV$\simeq $ $O(1000)$K below the Fermi level. 
Therefore, it seems unlikely that multiorbital effects affect the 
thermopower at least for $T\sim O(100)$K, where a large thermopower of 
$S\sim O(100)\mu$V/K is already observed.\cite{Lee} 
On the other hand, conventional calculations of the thermopower 
using the LDA bands 
already give large values,\cite{Singh,Wilson} 
suggesting that the band structure, not 
the electron correlation, is the key.
In fact, it has been pointed out in some studies  
that the narrowness of the $t_{2g}$ bands is essential.
\cite{Singh,Wilson,Takeuchi2} 
Although the narrowness of the bands must indeed be a factor,
this {\it alone} cannot be the origin of the large thermopower 
because otherwise we would expect more good thermoelectric materials.
Also, the problem of how the large thermopower and the large conductivity
can coexist has not been addressed clearly so far. 

\begin{figure}
\begin{center}
\includegraphics[width=8.7cm,clip]{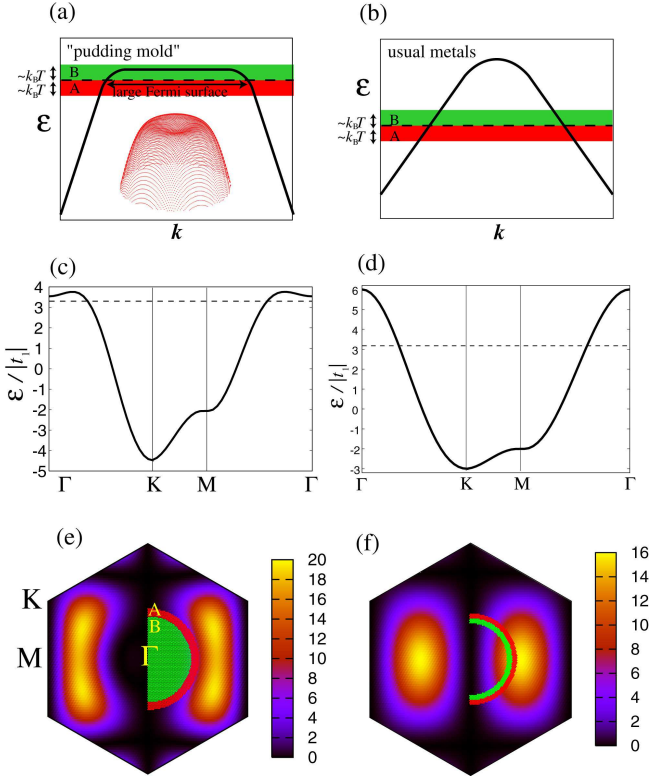}
\caption{(color online) 
Schematic figure for (a) the pudding mold band and 
(b) a usual metal. The inset of (a) shows the top of the band 
shown in (c). 
The band dispersion for (c) $t_2/t_1=-0.3$, $t_3/t_1=-0.11$ and 
(d) $t_2=t_3=0$. 
The dashed line shows the position of $\mu$ at $x=0.7$($n=1.7$) and $T=300$K.
A contour plot of $[v_x(\Vec{k})]^2$ for $x=0.7$ and $T=300$K with (e)
 $t_2/t_1=-0.3$, $t_3/t_1=-0.11$ and (f) $t_2=t_3=0$. 
Also region A (red outer half ring) and 
region B (green inner half circle or ring) are superposed. Only the 
$k_x\geq 0$ part of regions A and B is shown for clarity.
\label{fig1}}
\end{center}
\end{figure}

In the present study, we generally 
propose that a peculiar {\it shape} of the band (Fig.\ref{fig1}(a)), 
like the $a_{1g}$ band of the cobaltates (Fig.\ref{fig1}(c)), 
can give large thermopower and 
large conductivity at the same time. 
In fact, in the preceding studies, 
we have proposed that the superconductivity and the magnetism in 
Na$_x$CoO$_2$ 
both originate from the peculiar shape of the $a_{1g}$ band,\cite{KKprb,KKprl}
so adding the present study reveals that all of these interesting 
phenomena share the same origin.

Our idea is the following.
Using the Boltzmann's equation 
approach, the thermopower is given as\cite{Ashcroft}
\begin{equation}
{\bf S}=\frac{1}{eT}{\bf K}_0^{-1}{\bf K}_1
\end{equation}
where $e(<0)$ is the electron charge, $T$ is the temperature, 
tensors ${\bf K}_0$ and ${\bf K}_1$ are given by
\begin{equation}
{\bf K}_n=\sum_{\Vec{k}}\tau(\Vec{k})\Vec{v}(\Vec{k})\Vec{v}(\Vec{k})
\left[-\frac{\partial f(\varepsilon)}
{\partial \varepsilon}(\Vec{k})\right]
(\varepsilon(\Vec{k})-\mu)^n.
\label{eq2}
\end{equation}
Here, $\varepsilon(\Vec{k})$ is the band dispersion, 
$\Vec{v}(\Vec{k})=\nabla_{\Vec{k}}\varepsilon(\Vec{k})$ is the 
group velocity, $\tau(\Vec{k})$ is the quasiparticle lifetime,  
$f(\varepsilon)$ is the Fermi distribution function,  
and $\mu$ is the chemical potential. 
Hereafter, we simply refer to $({\bf K}_n)_{xx}$ as $K_n$, and 
$S_{xx}=(1/eT)\dot(K_1/K_0)$ (for diagonal ${\bf K}_0$) as $S$. 
Using $K_0$, conductivity can be given as 
$\sigma_{xx}=e^2K_0\equiv\sigma={1/\rho}$.
Roughly speaking for a constant $\tau$, 
\begin{eqnarray}
K_0\sim\Sigma'(v_A^2+v_B^2),\qquad
K_1\sim(k_BT)\Sigma'(v_B^2-v_A^2)
\end{eqnarray}
(apart from a constant factor) stand, 
where $\Sigma'$ is a summation over the states in the range of 
$|\varepsilon(\Vec{k})-\mu|<\sim k_BT$, 
and $v_A$ and $v_B$ are typical velocities for the states 
above and below $\mu$, respectively.
In usual metals, where $v_A\sim v_B$, 
the positive and the negative 
contributions in $K_1$ nearly cancel out to  
result in a small $S$ (Fig.\ref{fig1}(b)).
Now, let us consider a band 
that has a somewhat flat portion at the top (or the bottom), 
which sharply bends into a highly dispersive portion below (above).
We will refer to this band structure as the 
``pudding mold'' type (Fig.\ref{fig1}(a)). 
For this type of band with $\mu$ sitting near the 
bending point, $v^2_A\gg v^2_B$ holds for high enough temperature, 
so that the cancellation in $K_1$ is less effective, 
resulting in $|K_1|\sim(k_BT)\Sigma'v_A^2$ and $K_0\sim\Sigma'v_A^2$, 
and thus large $|S|\sim O(k_B/|e|)\sim O(100)\mu$V/K. 
%
A similar situation where a large $S$ originates from $v^2_A\gg v^2_B$ 
can be realized for a simple parabolic 
band if $\mu$ lies very close to the band edge,
but in that case, although the relative ratio $|\frac{K_1}{K_0}|$ is large,
$K_0$ is small because $v_A$, $v_B$ 
and the Fermi surface are small, 
so that the conductivity is small.
By contrast, for the pudding mold band,  
the large $v_A$ and  the large 
Fermi surface (in 2D and 3D) result not only in large  
$|\frac{K_1}{K_0}|\propto S$ but also in large $K_0\propto\sigma$ as well, 
being able to give a large power factor $S^2/\rho$, which is important
for device applications.\cite{Mahan} 
\begin{figure}[b]
\begin{center}
\includegraphics[width=8.7cm,clip]{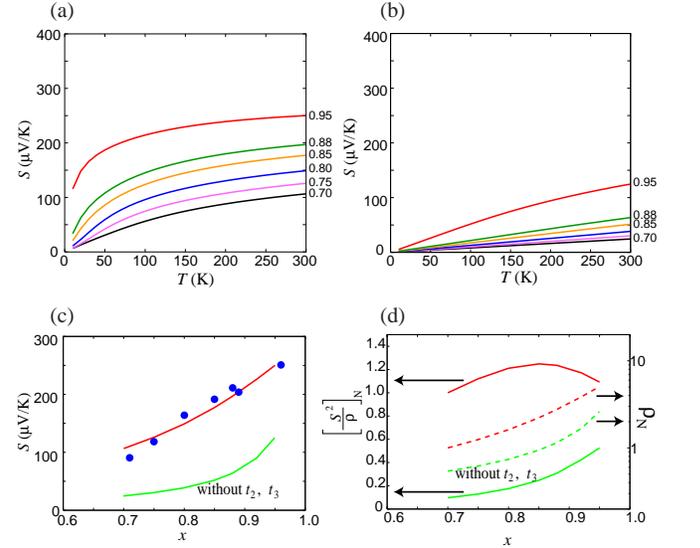}
\caption{(color online) 
(a) Thermopower for $t_2/t_1=-0.3$ and $t_3/t_1=-0.11$ 
plotted as functions of temperature for various $x$ 
(denoted at the right edge). 
$\Vec{k}$ dependence of $\tau$ is not considered here.
(b) Similar to (a) except $t_2=t_3=0$. 
(c) Thermopower and (d) 
the normalized (normalized by the values for $x=0.7$, 
$t_2/t_1=-0.3$ and $t_3/t_1=-0.11$ ) 
resistivity (dashed) and power factor (solid) at $T=300$K 
plotted as functions of $x$ for $t_2/t_1=-0.3$ and $t_3/t_1=-0.11$ (red) and 
$t_2=t_3=0$ (green). In (c), the experimental data at 300K 
(blue dots) are taken from ref.\cite{Lee}.
\label{fig2}}
\end{center}
\end{figure}

We now apply the idea to Na$_x$CoO$_2$. 
We take a 2D single band tight binding model
$H=-\sum_{ij}t_{ij}(c_i^\dagger c_j+c_j^\dagger c_i)$ 
that considers hopping integrals between 
first, second and third nearest neighbor sites, 
$t_1$, $t_2$, and $t_3$. $t_2/t_1=-0.3$, $t_3/t_1=-0.11$ 
are introduced so as to reproduce the shape of the 
$a_{1g}$ portion of the bands 
in the first principles calculation.\cite{Singh,Kunes} 
We take $t_1=-1000$K to reproduce the band width obtained 
in the ARPES data,\cite{Takeuchi,Yang2} 
which about 60$\%$ of the LDA result\cite{Singh,Kunes} 
due to electron correlation. The band filling $n$ is 
defined as $n=$number of electrons/sites, 
and it is related to the sodium content $x$ as $n-1=x$  
(as far as the $e_g'$ bands are fully filled).
In Fig.\ref{fig1}(c), we show the band dispersion,  
whose top around the $\Gamma$ point (inset of Fig.\ref{fig1}(a)) indeed 
has a form of a pudding mold.

In the calculation of $S$, we first neglect the $\Vec{k}$ 
dependence of $\tau$ (so that $\tau$ cancels out in $K_1/K_0$).
In Fig.\ref{fig2}(a), we show the temperature dependence of $S$ 
for various values of band fillings. 
If we first focus on $x=0.7$, we find that the overall temperature 
dependence observed experimentally for $x\simeq 0.71$ 
in ref.\onlinecite{Lee} as well 
as its large value of $\simeq 100\mu$V/K at $T\simeq 300$K
is reproduced. 
In Fig.\ref{fig2}(c), we show the Na content 
dependence of $S$ at $T=300$K, together with the experimental 
data of $S$ at 300K taken from ref.\onlinecite{Lee}.
Here again, we find a good agreement between 
the calculation and the experimental results. 
For comparison, we show in Figs.\ref{fig2}(b) and (c) 
the calculation result for 
$t_2=t_3=0$, where the band top is just parabolic (Fig.\ref{fig1}(d)). 
We see that the thermopower is strongly suppressed compared to the case with 
the pudding mold band. Note that for 
$t_2=t_3=0$, the band width itself is nearly the same with the case of 
$t_2/t_1=-0.30$ and $t_3/t_1=-0.11$, which indicates that 
the {\it shape} of the band is important.

In order to see that our idea is working, 
we show in the bottom of Fig.\ref{fig1} $[v_x(\Vec{k})]^2$, 
along with the regions 
$-2k_BT<\varepsilon(\Vec{k})-\mu<0$ (region A)
and $0<\varepsilon(\Vec{k})-\mu<2k_BT$ (region B).
It can be seen that for the pudding mold band in (e), 
$[v_x(\Vec{k})]_B^2/[v_x(\Vec{k})]_A^2\ll 1$, 
while for a parabolic band in (f) 
($t_2=t_3=0$), $[v_x(\Vec{k})]_B^2/[v_x(\Vec{k})]_A^2\simeq 1$. 
It is not appropriate 
to say that the large $S$ is due to the large DOS 
that originates from the flatness 
of the band top, because it is the {\it dispersive portion} (region A) 
that is positively contributing to $S$.

The important expectation for the pudding mold band 
is not just the large thermopower, 
but also a relatively large conductivity and thus a large power factor. 
In Fig.\ref{fig2}(d), we show the normalized resistivity $\rho_N$ and the 
power factor  $[S^2/\rho]_N$ with and without 
$t_2$ and $t_3$ (here the $x$ dependence of $\tau$ is 
neglected). The resistivity of the pudding mold band 
is only about a factor of $\sim 2$ larger than that for $t_2=t_3=0$, and 
this combined with the large thermopower 
results in a strong enhancement (10 times larger at $x=0.7$) 
of the power factor.\cite{comment}
\begin{figure}[b]
\begin{center}
\includegraphics[width=8.7cm,clip]{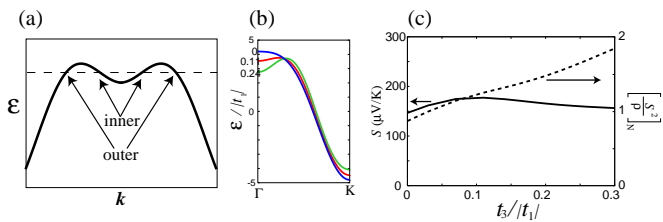}
\caption{(color online) 
(a) Schematic plot of the pudding mold band with corrugation.
$\mu$ sitting at the dashed line results in inner and outer Fermi surfaces.
(b) The variation of the band 
upon increasing $t_3/|t_1|$ as $0\rightarrow 0.11\rightarrow 0.24$ 
with $t_2/t_1=-0.3$.
(c) Thermopower (solid) and 
normalized power factor (dashed) plotted as functions of $t_3/|t_1|$ 
for $t_2/t_1=-0.3$, $x=0.85$ and $T=300$K. 
\label{fig3}}
\end{center}
\end{figure}

So far, we have not focused on the effect of multiple Fermi surfaces 
(Fig.\ref{fig3}(a)), 
which has been proposed to be the origin of the 
superconductivity\cite{KKprb,MO1} 
and the magnetism\cite{KKprl,Korshunov1,Hiroi}.
To see this effect,
we plot in Fig.\ref{fig3}(c) the $t_3$ dependence of 
$S$ and $[S^2/\rho]_N$. 
Increasing $t_3$ (or $t_2$) makes the 
local minimum structure at the $\Gamma$ point 
deeper as shown in Fig.\ref{fig3}(b).
$S$ takes its maximum at a certain $t_3$, 
but the power factor 
continues to grow. This is because 
there are inner and outer Fermi surfaces both 
contributing to the conductivity, whose increase overcomes the decrease of $S$.
If we go back to Fig.\ref{fig2}(d), this multiple Fermi surface 
effect is contributing to the low resistivity and high power factor for 
$x\sim 0.8$.

In this context, 
it is interesting to estimate how large a power factor 
the multiple $e_g'$ hole pockets would give if they were present. 
In fact, the top of the $e_g'$ bands around the K 
point is again of the pudding mold type, with some corrugation 
that results in the multiple pockets. 
If we use the effective single band dispersion that resembles the 
upper portion of the $e_g'$ bands\cite{KKprleg},
we find that the power factor is about two times larger than in the 
$a_{1g}$ case shown in Fig.\ref{fig2}(d). 
Thus, if it were possible to push up the $e_g'$ bands so that 
the band top comes above the Fermi level for high Na content,
the combined effect of both $a_{1g}$ and $e_g'$ pudding mold 
bands may result in a very large power factor.
Apart from this possibility, 
at high temperatures, and for relatively small Na content, 
where the Fermi level should sit close to the $e_g'$ band top, 
holes may be introduced at the top of the $e_g'$ band, which may be 
related to the enhancement of the thermopower above 500K 
observed in ref.\cite{Fujita}

\begin{figure}
\begin{center}
\includegraphics[width=8.7cm,clip]{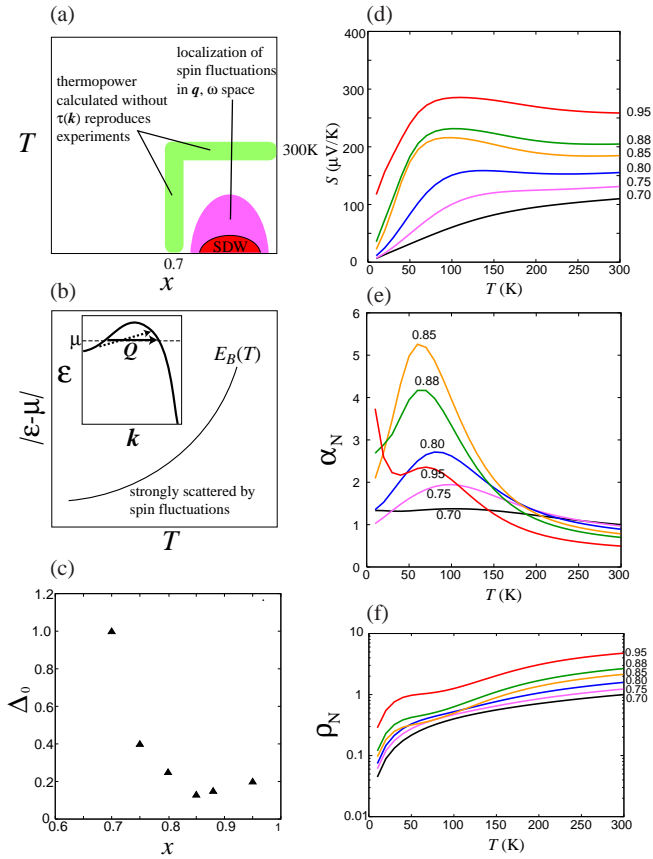}
\caption{(color online) (a) Schematic phase diagram concerning the 
relation between SDW and the thermopower. 
(b) The inset shows that when the spin fluctuations are localized in 
$\Vec{q},\omega$ space, 
the quasiparticle scatterings by spin fluctuations occur   
mainly with momentum transfer close to the nesting vector $\Vec{Q}$ 
and with low energy transfer (solid arrow), 
while scattering with other $\Vec{q}$ and large $\omega$ 
(dashed arrow) is less effective.
Shown in the main panel is a schematic plot of the energy scale $E_B(T)$ 
under which strong scatterings take place.
(c) Variation of $\Delta_0$ with $x$.
Calculation result for (d) the thermopower $S$, 
(e) the normalized Peltier conductivity $\alpha_N$ 
and (f) the normalized resistivity $\rho_N$ 
(normalized by the values at $x=0.7$ and $T=300$K)
for various $x$ using the trial form for $\tau$ 
with $\gamma=2$, $\xi_1=1$ and  
$\xi_2=0.1$ (in units of $|t_1|$).
\label{fig4}}
\end{center}
\end{figure}

We have seen that the consideration of the peculiar shape of the 
$a_{1g}$ band alone suffices to understand 
the experimentally observed thermopower for 
$x\simeq 0.7$ or $T\simeq 300$K.
For $x\geq 0.8$ and $T<200$K, however, this is not the whole story. 
In the calculated thermopower in Fig.\ref{fig2}(a), 
a hump structure around 100K found in the 
experiment for $x\geq 0.8$\cite{Lee} is not reproduced.
In fact, this Na content and the temperature range is close to the region 
where the metallic magnetism, or the spin density wave (SDW), 
\cite{Motohashi} sets in (Fig.\ref{fig4}(a)),
whose origin has been shown to be due to the nesting between the 
inner and the outer portions of the Fermi surface.
\cite{KKprl,Korshunov1}
When the system is close to the  SDW instability,  
the spin fluctuations 
become localized in $\Vec{q}$ space around the 
nesting vector $\Vec{Q}$ and in energy around $\omega\sim 0$.
Thus, the quasiparticle scatterings by spin fluctuations occur with 
momentum transfer close to $\Vec{Q}$ and 
with small energy transfer, so that, assuming a nearly isotropic 
Fermi surface, only the quasiparticles with 
$|\varepsilon(\Vec{k})-\mu|<E_B(T)$ is strongly scattered,
where $E_B(T)$ is a certain energy scale that decreases with 
temperature (Fig.\ref{fig4}(b)).
We leave microscopic treatment of this effect for future study,\cite{comment3} 
but here, in order to qualitatively take this effect into account, 
we assume a trial form of $\tau$ as 
$1/\tau(\Vec{k},T)\propto 
A^2(\Vec{k},T)T+[1-A^2(\Vec{k},T)]T^2/\xi_1$, 
$A(\Vec{k},T)=\cosh^{-1}\{[\varepsilon(\Vec{k})-\mu(T)]/E_B(T)\}$.
This form assumes that $\tau$ in the strongly scattered regime is 
proportional to $1/T$, while it crosses over to $1/T^2$ in the 
weakly scattered regime. For $E_B(T)$, we assume a form 
$E_B(T)=k_BT(T/\xi_2)^\gamma+\Delta_0(x)$, where  
$\xi_2$ is the temperature scale below which $E_B(T)$,
apart from $\Delta_0$, becomes smaller than the energy scale $(\sim k_BT)$ 
relevant for $K_0$ and $K_1$. 
We vary $\Delta_0$ with $x$ as shown in Fig.\ref{fig4}(c) 
so as to reflect the tendency towards the SDW formation, which is 
the strongest around $x=0.8\sim 0.85$ experimentally.
$S$, the Peltier conductivity $\alpha=S\sigma$, 
and $\rho$ (Fig.\ref{fig4}(d)-(f)) calculated by using this $\tau$
and taking $\gamma=2$, $\xi_1=1$, and  
$\xi_2=0.1$ (in units of $|t_1|$)  
indeed reproduce the experimental 
results at least qualitatively.
At low temperatures for $x\geq 0.75$,  $\alpha$ 
exhibits a maximum and $\rho$ crosses 
over from $T$ to $T^2$ dependence 
in rough agreement with the experiment.\cite{Lee} 
This is because the quasiparticles somewhat 
away from the Fermi surface have enhanced lifetime at low 
temperatures.\cite{comment2}
The thermopower $S=\alpha/\sigma$  
now has the hump structure at low temperatures for $x\geq 0.8$ because 
$K_1\propto\alpha T$ is enhanced more than $K_0\propto \sigma$ 
since the former 
has larger contribution from the states away from the Fermi surface 
due to the factor $(\varepsilon(\Vec{k})-\mu)$ in eq.(\ref{eq2}).
At present, our form of $\tau$ has no 
microscopic basis, so we do not persist in these particular 
parameter values. 
In fact, for $x=0.85$, e.g., by taking $\gamma=1$, $\xi_2=0.06$,
$\Delta_0=0.08$, a similar curve for $S$ 
but with a smaller hump is obtained. 

To summarize, we have pinned down the origin of the 
coexistence of the large thermopower and the large conductivity in 
Na$_x$CoO$_2$: 
the pudding mold $a_{1g}$ band that has a dispersive portion and 
a flat portion with some corrugations at the top that can result in 
multiple Fermi surfaces. The present study provides a new guiding 
principle for designing good thermoelectric materials: look for 
materials that have  
pudding mold band(s) with the chemical potential 
lying close to its bending point.
Roughly speaking, pudding mold bands tend to 
occur in geometrically frustrated lattice structures because 
there are several hopping contributions to $\varepsilon(\Vec{k})$, 
which nearly cancel at some $\Vec{k}$, while not at other $\Vec{k}$.
This, however, does not mean that ``seemingly nonfrustrated'' 
lattices are always not good  since, for example, 
a square lattice with 
sufficiently large diagonal hopping integrals $(t')$ in addition to 
vertical and horizontal hoppings $(t)$ 
also results in a pudding mold band and a large thermopower 
(e.g., for $t=-1$ and $t'=0.45$, $S\simeq 200\mu$V/K at 
$T=0.3|t|$ and $n=1.80$).
Another caution is that, even if a pudding mold band is present,
a coexistence of a dispersive metallic 
band with a much larger $K_0$ and a small $K_1$ would 
result in a suppression of 
$(\sum_{bands}K^{band}_1)/(\sum_{bands}K^{band}_0)$ and thus a small total 
thermopower. Therefore, reliable band structure 
calculation is necessary in order to design good thermoelectric 
materials based on the present guiding principle.
%

Numerical calculations were performed
at the facilities of the Supercomputer Center,
ISSP, University of Tokyo.
This study has been supported by 
Grants-in-Aid for Scientific Research from the Ministry of Education, 
Culture, Sports, Science and Technology of Japan, and from the Japan 
Society for the Promotion of Science.

%



\begin{thebibliography}{99}
\bibitem{comment} The origin of the strong enhancement of the power factor 
differs from the common 
knowledge that a parabolic band with 
large effective mass $m^*$ 
gives a large thermopower.  For a fixed carrier density, 
the thermopower for the parabolic band 
is proportional to $m^*$,\cite{Ashcroft} 
but this can be attributed solely to 
$K_0\propto\sigma\propto 1/m^*$, while $K_1$ is independent
of $m^*$. Thus, the power factor $\propto K_1^2/K_0$ 
is only proportional to the resistivity.
By contrast, in the pudding mold case, 
the flat portion of the band contributes to an  
increase of $K_1$, so that 
an increase of only a factor of 2 in the resistivity can result in 
a power factor 10 times larger.
\bibitem{comment3} A microscopic investigation on this issue may require a 3D 
model, since the actual Fermi surface 
nesting is 3D.\cite{KKprl}
\bibitem{comment2} The deviation of $\rho$ and $\alpha$ from the experiments 
below 50K is due to the  divergence of $\tau(\Vec{k})$ 
at very low temperatures, which should not occur in actual materials.
\end{thebibliography}
\end{document}